    \documentstyle[aps, preprint]{revtex}
    \tightenlines
    
    \def\be{\begin{equation}}
    \def\ee{\end{equation}}
    \def\ba{\begin{eqnarray}}
    \def\ea{\end{eqnarray}}

    \input epsf
    \begin{document}
    \title{Energy-Momentum Tensor of Field Fluctuations in Massive
    Chaotic Inflation}
    \author{F. Finelli \footnote{e-mail: fabio@physics.purdue.edu \\
                 $ \hspace*{0.2cm}^\dag \!\!$ e-mail: marozzi@bo.infn.it\\
                 $\hspace*{0.2cm}^\ddag \!\!$ e-mail: vacca@bo.infn.it \\
                 $\hspace*{0.2cm}^\S \!\!$ e-mail: armitage@bo.infn.it }
    $^{\,1,2}$,
    G. Marozzi$^\dag \ ^{\,2}$,
    G. P. Vacca$^\ddag \ ^{\,3,2}$
    and G. Venturi$^\S \ ^{\,2}$ }
    \address{$^1$ Department of Physics, Purdue University, West Lafayette,
    IN 47907, USA \\
    $^2$ Dipartimento di Fisica, Universit\`a degli Studi di Bologna
    and I.N.F.N., \\ via Irnerio, 46 -- 40126 Bologna -- Italy \\
    $^3$ H. Institut f\"ur Theoretische Physik, Universit\"at Hamburg, Luruper
    Chaussee 149, \\ D-33761 Hamburg, Germany}
    \maketitle
    \begin{abstract}
    We study the renormalized energy-momentum tensor (EMT) of the inflaton
    fluctuations in rigid space-times during the slow-rollover regime for
    chaotic inflation with a mass term.
    We use dimensional regularization with
    adiabatic subtraction
    and introduce a novel analytic approximation for the inflaton
    fluctuations which is valid during the slow-rollover regime.
Using this approximation we find a scale invariant spectrum for the inflaton 
fluctuations in a rigid space-time, and we confirm this result by numerical 
methods. The resulting renormalized EMT is covariantly
    conserved and agrees with the Allen-Folacci result in the de Sitter limit,
    when the expansion is exactly linearly exponential in time.
    We analytically show that the EMT tensor of the
    inflaton fluctuations grows initially in time, but saturates to the value
    $H^2 H^2_0$, where $H$ is the Hubble parameter and $H_0$ is its 
    value when inflation has started. This result also implies that
    the quantum production of light scalar fields (with mass smaller or 
equal to the
    inflaton mass) in this model of chaotic inflation depends on the duration
    of inflation and is larger than the usual result
    extrapolated from the de Sitter result.

    \end{abstract}

    \section{Introduction}

    Particle production in expanding universe, pionereed by L. Parker
    \cite{parker1}, is an essential ingredient of inflationary cosmology
    \cite{BOOKS}. The nearly scale invariant spectrum of density
    perturbations predicted by inflationary models \cite{all} is at present
    inextricably related
    to the concept of amplification of vacuum fluctuations by the geometry.
    The scale invariant
    spectrum for massless minimally coupled fields during a de Sitter era was
    indeed computed before inflation was suggested.

    The calculation of the energy carried by these amplified
    fluctuations is then {\em the} natural question. To answer this question a
    renormalization scheme is necessary, as in ordinary Minkowski 
space-time.
    Ultraviolet divergences due to
    fluctuations on arbitrary short scales are common in field theory.
    In Minkowski space-time,
    infinities in a free theory are removed by the
    subtraction of the vacuum expectation value of the energy, also called
    normal ordering, the physical justification being that these vacuum
    contributions are unobservable.

    A similar prescription is used in order to regularize infinities in
    cosmological space-times. However, one additional problem is the absence
    of an unambiguous choice of vacuum, because of the absence of a class of
    privileged observers, which are the inertial observers in the 
Minkowski
    space-times. The idea is then to subtract the energy associated with a
    vacuum for which the effects of particle production by the
    time-dependence of the metric are minimized. This vacuum is determined
    by the assumption of an adiabatic expansion of the metric. This procedure
    is therefore called {\em adiabatic subtraction} \cite{adiabatic,birrell}.
    In this way the infinities of field theory are removed.

    Even with a minimal prescription such as the above,
    there are several surprising
    effects accompanying the renormalized energy-momentum tensor (henceforth
    EMT) of a test field in cosmological space-times.
    To name a few, there could be the avoidance
    of singularities due to quantum effects \cite{parkerfulling},
    conformal anomalies which break classical conformal symmetries at the
    quantum level \cite{duff}, violation of the various energy conditions
    \cite{visser}. One of the first models of inflation proposed by
    Starobinsky \cite{rsquare} was indeed based on the role of the conformal
    anomaly, which both avoids the singularity and
    produces an inflationary phase.

    While the adiabatic vacuum for a test field - and its associated EMT - can
    be computed for generic cosmological space-times, the unrenormalized EMT
    can be calculated analytically only if exact analytic solutions for the
    field Fourier modes are available.
    Because of this, space-times such as de Sitter have been
    throughly investigated \cite{ren_desitter,birrell}, since analytic
    solutions for a scalar field with
    generic mass and coupling are available. In the absence of analytic
    solutions, numerical schemes are implemented \cite{numerical}.
    In a de Sitter space-time,
    the back-reaction of a test field seems important only if $m^2+\xi R = 0$,
    with $m$ and $\xi$ separately different from zero, or $m^2 + \xi R < 0$
    ($m$ is the mass of the test field and $\xi$ its coupling to the
    curvature $R$) \cite{HMPM,attractor}. In the former case, the EMT of the
    test field grows linearly in time, while in the latter case it grows
    exponentially. The EMT of a massless minimally coupled field is constant
    in de Sitter space-time \cite{AF}.

    Inflationary models based on the use of scalar
    fields, have an accelerated stage, usually called {\em slow rollover}, in
    which the Hubble parameter is almost frozen.
    During this stage it is rare to have exact
    solutions for the fluctuations.
    However, these inflationary models are more attractive
    than the de Sitter space-time one in furnishing phenomenological models
    whose predictions can be tested against observations. Chaotic models
    \cite{chaoticlinde} are the simplest among these. In this paper we study
    {\em analytically} the problem of back-reaction of inflaton
    fluctuations during the regime of slow-rollover for the case of a massive
    inflaton.  We consider inflaton fluctuations in rigid space-times, i.e. we
    neglect metric perturbations coupled to them, as pioneeringly investigated
    by Abramo, Brandenberger and Mukhanov \cite{ABM}. We plan to come back to
    this issue in a future work \cite{new}.

    The plan of the paper is as follows: in section II we describe the
    background classical dynamics for a massive inflaton and the novel
    analytical approximation for its fluctuations during the
    slow-rollover regime. In section III we discuss the normalization of
    quantum fluctuations and in section IV we compare the numerical evaluation
    of the spectrum with the analytic approximation. We discuss the
    EMT of inflaton fluctuations, the adiabatic subtraction and the
    renormalization in sections V-VII, respectively. In section VIII the
    problem of the back-reaction of the EMT of inflaton fluctuations is
    addressed
    and in section IX our novel analytic approximation is compared with the
    slow-rollover technique \cite{SL}. In section X we analyze the 
    production of a secondary massive field $\chi$ lighter than the 
    inflaton and we show that its production depends on the duration of 
    inflation. In section XI we conclude and in the two
    appendices we relegate useful formulae for the adiabatic expansion and the
    dimensional regularization with cut-off.


    \section{Analytic approximation}

    We consider inflation driven by a classical minimally coupled
    massive scalar field. The action is:
    \be
    S \equiv \int d^4x {\cal L} = \int d^4x \sqrt{-g} \left[ \frac{R}{16{\pi}G}
    - \frac{1}{2} g^{\mu \nu}
    \partial_{\mu} \phi \partial_{\nu} \phi
    - \frac{1}{2} m^2 \phi ^2 \right] \,
    \label{action}
    \ee
    where ${\cal L}$ is the lagrangian density and $m$ is the mass
    of the field $\phi$.
    Further we consider the Robertson-Walker line element with flat spatial
    section:

    \be
    ds^2 = g_{\mu \nu} dx^{\mu} dx^{\nu} = - dt^2 + a^2(t) d\vec{x}^2
    \label{metric}
    \ee

\begin{figure}
    \leavevmode\epsfysize=2.5in \epsfbox{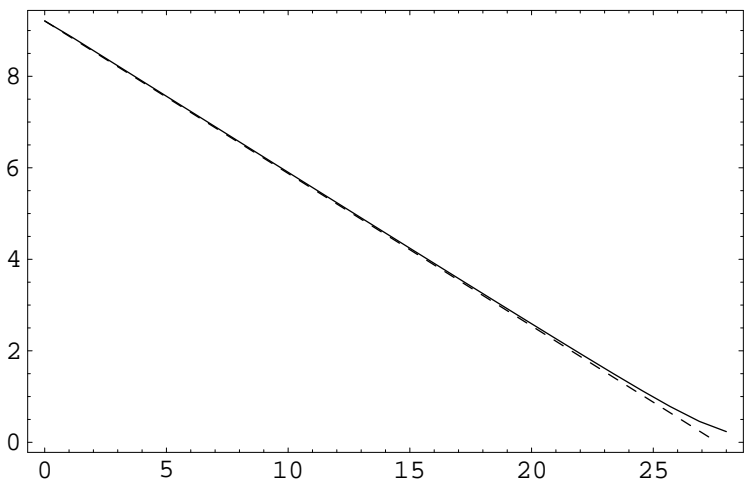}
    \caption{We plot the numerical evolution of $H$ (solid line) and the
    analytic approximation (dashed line). The initial condition corresponds to
    $H(t=0)\approx 9.2  m$ and the time is in $1/m$ units.}
    \label{fig:hubble}
    \end{figure}

    The scalar field is separated in its homogeneous component
    and the fluctuations around it, $\phi(t, {\bf x}) = \phi(t) +
    \varphi (t, {\bf x})$. During slow rollover the potential energy dominates
    and therefore
    \be
    H^2 = \frac{\kappa^2}{3} \left[ \frac{\dot\phi^2}{2} + m^2
    \frac{\phi^2}{2} \right] \simeq \kappa^2 \frac{m^2 \phi^2}{6}
    \label{hubble}
    \ee
    where $\kappa^2 = 8 \pi G = 8 \pi/M_{\rm pl}^2$.
    The Hubble parameter evolves as:
    \be
    \dot H = - \frac{\kappa^2}{2} \dot \phi^2
    \label{hubbleder}
    \ee
    On using the equation of motion for the scalar field:
    \be
    \ddot \phi + 3 H \dot \phi + m^2 \phi = 0
    \label{scalar_hom}
    \ee
    and neglecting the second derivative with respect to time we have $3 H
    \dot \phi \simeq - m^2 \phi$, and from
    Eq. (\ref{hubbleder}) one obtains:
    \be
    \dot H \simeq - \frac{m^2}{3} \equiv \dot H_0
    \label{hdot}
    \ee
    which leads to a linearly decreasing Hubble parameter and correspondingly
    to an evolution for the scale factor which is not exponentially linear in
    time, i.e.:
    \be
    H(t) \simeq H_0 + \dot H_0 t \quad a(t) \simeq \exp[ H_0 t + \dot H_0
    \frac{t^2}{2} ]
    \label{evolution}
    \ee
    In Fig. \ref{fig:hubble} the comparison of the analytic
    approximation with the numerical evolution for the Hubble parameter $H$ is
    displayed.

    We now consider the equations of motion for the inflaton fluctuations
    $\varphi$ in a rigid space-time (i.e. without metric perturbations):

    \be
    \ddot{\varphi}_{\bf k} + 3 H \dot{\varphi}_{\bf k} +
    \omega_k^2 \varphi_{\bf k} = 0 \,,
    \label{scalar1}
    \ee
    where $\omega_k^2 = k^2/a^2 + m^2$ and the
    $\varphi_{\bf k}$ are the Fourier modes of the inflaton fluctuations,
    \be
    \varphi (t, {\bf x}) = \frac{1}{\sqrt{V}} \sum_{\bf k}
    \left[ e^{i {\bf k} \cdot {\bf x}} \varphi_{k} (t) +
    e^{- i {\bf k} \cdot {\bf x}} \varphi^*_{k} (t) \right]
    \label{scalarFourier}
    \ee

    As already emphasized in the introduction, exact solutions for
    scalar fields in an expanding universe are rare,
    and indeed we do not have exact solutions for Eq. (\ref{scalar1}) with the
    time evolution given by Eq. (\ref{evolution}). We therefore
    introduce an approximation scheme based on an analogy with de Sitter
    space-time, where
    exact solution for scalar fields with arbitrary mass and coupling to the
    curvature do exist.
    We introduce $\psi_{\bf k} = a^{3/2} \varphi_{\bf k}$ and we split the
    time dependence in $\psi_{\bf k}$ as follows:
    \be
    \psi_{\bf k} = \psi_{\bf k} (\zeta, H) \quad {\rm with} \quad \zeta =
    \frac{k}{a H}  \,.
    \label{zdef}
    \ee
    The equation for $\psi_{\bf k}$ is
    \be
    \ddot \psi_{\bf k} + \left[ \frac{k^2}{a^2} + m^2 -\frac{3}{2} \dot H -
    \frac{9}{4} H^2 \right] \psi_{\bf k} = 0 \,.
    \label{psi}
    \ee
    We now make the ansatz $\psi_{\bf k} = \zeta^{\mu} Z_\nu (\lambda \zeta)$
    with $\mu$, $\nu$ and $\lambda$ functions of $H$. On expressing the first
    and second time derivatives as derivatives
    with respect to $(\zeta,H)$ and using $\ddot H \simeq 0$, as follows from
    Eq. (\ref{hdot}), we obtain from Eq. (\ref{psi}) after a little algebra:
    \begin{eqnarray}
    \zeta^\mu && \left[ \zeta^2 \frac{\partial^2}{\partial \zeta^2} +
    \zeta
    \frac{\partial}{\partial \zeta} + (\lambda^2 \zeta^2 - \nu^2 ) \right]
    Z_\nu + \nonumber \\
    &&
    {\rm Res}_1 \, \zeta^{\mu +1} \, \frac{\partial Z_\nu}{\partial \zeta} +
    {\rm Res}_2 \, \zeta^{\mu+2} \, Z_\nu + {\rm Res}_3  \,
    \zeta^{\mu} \, Z_\nu
    + {\cal O} (\frac{\dot H^2}{H^4}) = 0  \,
\label{maineq}    
\end{eqnarray}
    where we have neglected quadratic and higher order terms in $\dot{H}/H^2$.
    Indeed, in order to have a value of density perturbations compatible
    with observations,
    $m$ is constrained to be ${\cal O} (10^{-5}-10^{-6} M_{\rm pl})$: from
    Eqs. (\ref{hubble},\ref{hdot}) one can
    see that working
    to first order in $\dot{H}/H^2$ during slow-rollover ($\phi \sim {\rm few}
    M_{\rm pl}$) is a good
    approximation.
    On considering $H$ and $\zeta$ as independent variables, the first term
    vanishes if $Z_\nu$ is a Bessel function of argument $\lambda \zeta$ and
    index $\nu$.
    On requiring that the residual functions ${\rm Res}_i , \, i=1-3$ vanish
    individually, the parameters $\lambda, \mu$ and $\nu$ are determined to
    be:
    \be
    \lambda = 1 - \frac{\dot H}{H^2} \,, \quad \mu = \frac{\dot H}{2 H^2}
    \label{parameters1}
    \ee
    \be
    \nu^2 = \frac{9}{4} - \frac{m^2}{H^2}
    - 3 \frac{\dot H}{H^2} \,.
    \label{parameters2}
    \ee
    Hence the general solution to Eq. (\ref{scalar1}) is:
    \be
    \varphi_{\bf k} = \frac{1}{a^{3/2}} \zeta^\mu \left[ A H_\nu^{(1)}
    (\lambda \zeta) + B H_\nu^{(2)} (\lambda \zeta) \right]
    \label{solution}
    \ee
    where $H_\nu^{(1,2)}$
    are the Hankel functions of first and
    second kind respectively, and $A, B$ are time-independent coefficients to
    the order of our approximation.

    We note that in the de Sitter limit ($\dot H/H^2 \rightarrow 0$) the
    solution in Eq. (\ref{solution}) tends to the de Sitter solution
    \cite{ren_desitter,sol_desitter,BD}, since $\lambda = 1 \,, \mu=0$. 

    On using Eq. (\ref{hdot}) the value for the index $\nu$ in Eq.
    (\ref{parameters2}) corresponds to
    an exact scale invariant spectrum for the inflaton fluctuations
    $\varphi$, i.e. $\nu = 3/2$.
    We shall show numerically in section IV that this analytic approximation
    is very good for the relevant spectrum range. This numerical analysis 
agrees with a previous numerical estimate of the same spectral index 
\cite{kuztka}. This scale invariant spectral index could seem a little 
surprising,
    since in de Sitter space-time, a mass term
    would lead to a spectrum, which is slightly blue shifted with respect to
    scale invariance ($\nu < 3/2$). To see this, it is sufficient to put
    $\dot H = 0$ in Eqs. (\ref{psi},\ref{parameters2}). On considering $\dot H
    \ne 0$ (and negative), it appears to give a positive contribution to the
    mass term
    in Eq. (\ref{psi}), instead it compensates the mass term in Eq.
    (\ref{parameters2}). The interpretation is the following: on considering a
    de Sitter stage in which $H$ slowly decreases, a fluctuation
    freezes when it crosses the Hubble radius, with an amplitude
    determined by the value of the Hubble radius at the horizon crossing.
    However $H$ decreases, therefore if $k_1 > k_2$, this effect implies that
    the amplitude for the
    mode $k_1$ is smaller than the one for the mode $k_2$, since the latter
    crosses the Hubble radius first. This effect is a
    red tilt of the de Sitter scale invariant spectrum. For the case of
    slow-rollover in a chaotic
    inflationary model with a massive inflaton, these red and blue
    shifts exactly compensate, leading to a scale-invariant spectrum for
    the inflaton fluctuations $\varphi$ in rigid space-time.

    \section{Quantized fluctuations}

    We now consider quantized fluctuations of the inflaton. This means that
    Eq. (\ref{scalarFourier}) is promoted to an operator form:
    \be
    \hat{\varphi} (t, {\bf x}) = \frac{1}{\sqrt{V}} \sum_{\bf k}
    \left[ \varphi_{k} (t) \, e^{i {\bf k} \cdot {\bf x}} \, \hat{b}_{\bf
    k} + e^{- i {\bf k} \cdot {\bf x}} \varphi^*_{k} (t)
    \hat{b}^\dagger_{{\bf k}} \right]
    \label{quantumFourier}
    \ee
    where the $\hat{b}_k$ are time-independent Heisenberg operators (also
    called time independent invariants in \cite{desitter,nonzero}).
    In order to have the usual commutation relations among the $\hat{b}_k$:
    \be
    [\hat{b}_{\bf k}, \hat{b}_{{\bf k}'}] =
    [\hat{b}^\dagger_{\bf k}, \hat{b}^\dagger_{{\bf k}'}] = 0 \, \quad
    [\hat{b}_{\bf k}, \hat{b}^{\dagger}_{{\bf k}'}] = \delta^{(3)}
    ({\bf k} - {\bf k}') \,
    \ee
    one must normalize the solution to the equations of
    motion through the Wronskian condition:
    \be
    \varphi_{\bf k} \, \dot{\varphi}^*_{\bf k} -
    \dot{\varphi}_{\bf k} \varphi^*_{\bf k} = \frac{i}{a^3} \,.
    \label{wronskian}
    \ee
    This normalization condition yields to the following relation among
    the coefficients $A, B$ of Eq. (\ref{solution}):
    \be
    |A|^2 - |B|^2 = \frac{\pi}{4 H} \, \lambda \, \zeta^{-2\mu} \,.
    \ee
    The fact that $A, B$ depend on time should
    not surprise. In time-dependent perturbation theory, these coefficients,
    which would be time independent for exact solutions, acquire a time
    dependence \cite{sakurai}, just as the Wronskian of the solutions. In our
    case, this time dependence is consistent with the approximation, i.e.
    self-consistent to (including) order $\dot H/H^2$.

    The solution corresponding to the Bunch-Davies vacuum \cite{BD} in
    de Sitter space-time, that is the adiabatic vacuum for $k \rightarrow \infty$
    during the slow-rollover phase,
    corresponds to choosing $A = (\pi \lambda/4H)^{1/2} \zeta^{-\mu} 
\,, B=0$.
    With this choice, for $\lambda \zeta \rightarrow \infty$,
    the solution (\ref{solution})
    becomes:
    \be
    \varphi_{\bf k} \simeq -\frac{1}{a \sqrt{2 k}}
    e^{+i \lambda \zeta}
    \label{adiabatic_limit}
    \ee
    Let us now discuss the behaviour for $\lambda \zeta \ll 1$.
    On using Eqs. (\ref{zdef},\ref{parameters1}) one
    sees that $\lambda \zeta \ll 1$ implies $k \ll a H$, i.e. wavelengths
    which are much larger than the Hubble radius. In this limit, the
    solution is \cite{AS}:
    \be
    \varphi_{\bf k} \simeq - i \frac{\Gamma(\nu)}{\pi \, a^{3/2}}
    \left( \frac{\pi \lambda}{4 H} \right)^{1/2} \left( \frac{\lambda \zeta}{2}
    \right)^{-\nu}.
    \label{longw}
    \ee
    In order to compute expectation values of operators with respect
    to states in the
    time-independent invariant $b$ \cite{desitter,nonzero} basis it is useful
    to introduce the modulus of mode functions $x_k =
    |\varphi_{\bf k}/\sqrt{2}|$.
    The variable $x_k$
    satisfies the following Pinney equation:
    \be
    \ddot x_k + 3 H \dot x_k + \omega_k^2 x_k = \frac{1}{a^{6} x_k^3} \,.
    \label{pinneyx}
    \ee
    We now rescale $x_k = \rho_k/a^{3/2}$ to eliminate the damping term and
    obtain:
    \be
    \ddot \rho_k + \left[ \omega_k^2 - \frac{9}{4} H^2 - \frac{3}{2} \dot H
    \right] \rho_k = \frac{1}{\rho_k^3} \,.
    \label{pinneyrho}
    \ee
    The general solution to Eq. (\ref{pinneyrho}) is given as a nonlinear
    combination of two independent solutions $y_1, y_2$ to the linear part of
    Eq. (\ref{pinneyrho}).
    From Eq. (\ref{psi}) we can use the following solutions:
    \begin{eqnarray}
    y_1 &=& \zeta^\mu J_\nu(\lambda \zeta) \nonumber \\
     y_2 &=& \zeta^\mu N_\nu(\lambda \zeta) \label{newsolution2B}
    \end{eqnarray}
    Therefore the solution to Eq. (\ref{pinneyrho}) is
    \be
    \rho_k = \left( L y_1^2 + M y_2^2 +2 N y_1 y_2 \right)^\frac{1}{2}
    \ee
    where
    the coefficients satisfy $LM - N^2 = 1/\bar{W}^2$, with $\bar{W}$ the
    (time-dependent) Wronskian of $y_1, y_2$. The choice of initial
    conditions for $\rho_k$ which
    corresponds to the adiabatic vacuum for $k \rightarrow \infty$
    is $N=0$ and $L=M=\lambda \pi/(2 H
    \zeta^{2\mu})$. The solution for $x_k$  is:
    \be
    x_k = \frac{1}{a^{3/2}} \left( \frac{\pi}{2 H} \right)^{1/2}
    \lambda^{1/2} \left[ J_\nu^2 (\lambda \zeta ) + N_\nu^2 (\lambda \zeta)
    \right]^{1/2} \,
    \label{xsolution}
    \ee
    which coincides with the Bunch-Davies choice in the de Sitter limit
     \cite{desitter}.
  
    \section{Numerical Analysis}

    In this section we present the numerical analysis of the time evolution of
    the $\varphi$ modes. Besides checking the validity of the analytical
    approximation introduced in section II, this analysis is useful in order
    to understand how a natural infrared cut-off emerges in the problem, on
    assuming that inflation is not eternal in the past, but starts at some
    finite time. This infrared cut-off becomes relevant when the $\varphi$
    fluctuations are generated in an infrared state \cite{fordparker}, as
    occurs for $\nu \ge 3/2$ (see Eq. (\ref{parameters2})).

First we want to analyze the properties
of the spectrum of the inflaton fluctuations.
    We numerically evolve Eqs. (\ref{hubble},\ref{scalar_hom}) and Eq.
    (\ref{pinneyx}).
    We present numerical data for an interval of comoving wavenumbers for
    which $1 \le k/m \le 10^5$ at the initial time $t_0$ ($a(t_0) = 1$).
    The initial conditions are $\phi(t_0)= 4.5 M_{\rm pl}$,
    $\dot \phi (t_0) = 0$ for the inflaton.
    If we consider the vacuum state for each mode of the field fluctuations
    to be the initial condition, one has

    \begin{eqnarray}
    x_k(t_0) &=& \frac{1}{a^{3/2}(t_0) \omega_k^{1/2}(t_0)}  \nonumber  \\
    \dot{x}_k(t_0) &=& 0
    \label{initpinney}
    \end{eqnarray}
    This fact can be easily seen in terms of the invariant operators introduced to
    quantize time dependent harmonic oscillators \cite{desitter,nonzero,chaotic}.

\begin{figure}
    \leavevmode\epsfysize=2.5in \epsfbox{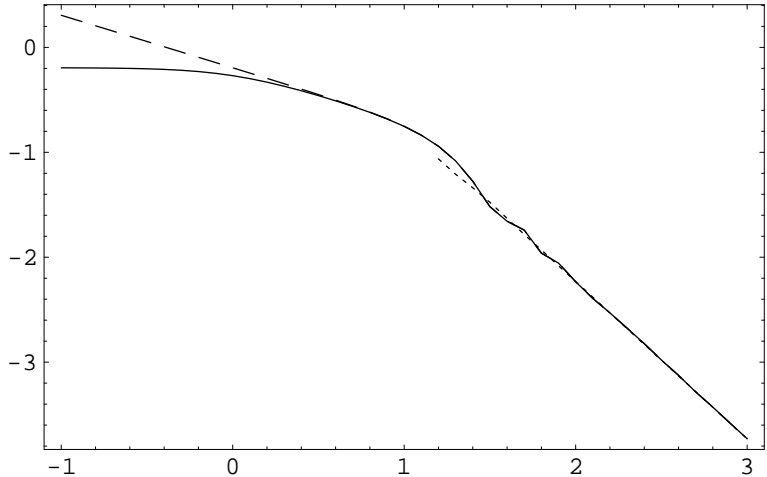}
    \caption{ We show on a logarithmic scale the r.m.s. of the inflaton
    fluctuations $x_{\bf k}$ obtained numerically for the three different sets
    of initial conditions. 
The spectrum is shown as function of ${\rm Log}(k/m)$. }
    \label{fig:comparison}
    \end{figure}

    As a second initial condition, we consider the limit for large $k$
    of the conditions in (\ref{initpinney}), which correspond to setting the
    mass equal
    to zero. A third set of initial conditions is related to the adiabatic
    expansion in conformal time, (see later Eq.  (\ref{conformal_eq})):
    \begin{eqnarray}
    x_k(t_0) &=& \frac{1}{a(t_0) \Omega_k^{1/2}(t_0)}   \nonumber \\
    \dot x_k(t_0) &=& -  H(t_0) x_k(t_0)
    \end{eqnarray}
    where $\Omega_k$ will be defined in Eq. (\ref{conformal_freq}).
    Let us note that for the last case the frequency becomes imaginary below a
    certain threshold and so we shall consider the region above it.
    In Fig. \ref{fig:comparison} we exhibit the three cases.
    The first two initial conditions lead to spectra practically equal
    for $k$ of order $m$ and above, the third set instead has a spectrum which
    joins the others at values of $k$ of the order of $H_0$.

\begin{figure}
    \leavevmode\epsfysize=2.5in \epsfbox{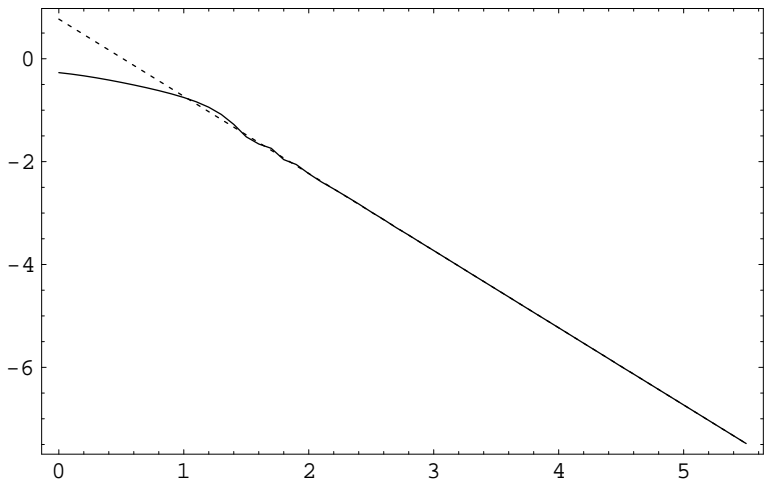}
    \caption{We show on a logarithmic scale the r.m.s. of the inflaton
    fluctuations $x_{\bf k}$ obtained numerically (solid line)
    and and an asymptotic linear fit (dotted line).
    The spectrum is shown as function of ${\rm Log} (k/m)$ and
    the linear fit is given by $0.7708 - 1.5 \,  {\rm Log} (k/m)$, in 
agreement with [21].}
    \label{fig:spettro}
    \end{figure}

    The spectrum of the fluctuations, related to the initial conditions in
    (\ref{initpinney}),
    are displayed, over a broader range, in Fig.  \ref{fig:spettro}.

    Figs. \ref{fig:comparison} and \ref{fig:spettro} display the
    spectrum at $t=10/m$ (for this case inflation lasts a period 
of time $\sim 27/m$).
    We note that this scale invariant spectrum extends only up to a
    certain scale, $\ell$, which is of the same order as the initial Hubble
    radius
    $H(t_0)$ (we have checked this by changing the initial conditions for the
    homogeneous mode of the inflaton). For comoving modes smaller than $\ell$
    the spectrum oscillates and
    bends towards the initial conformal adiabatic vacuum state, as shown in
    Fig. (\ref{fig:spettro}).
    For all practical purposes we can therefore safely consider a scale invariant
    spectrum for $k > l$, where $\ell = C H (t_0)$,
    with $C$ a numerical coefficient ${\cal O} (1)$.
    This numerical evidence
    favours the picture in which the amplification of the modes occurs
    mainly at the crossing of Hubble radius. Since $m << H$ during
    inflation,
    this means that all the modes for which $m \le k \lesssim H(t_0)$, are not
    stretched by the geometry.

\begin{figure}
    \leavevmode\epsfysize=2.5in \epsfbox{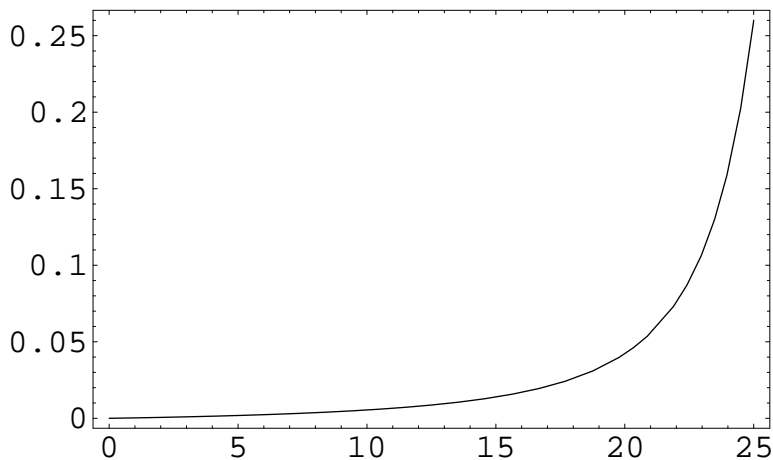}
    \caption{ Relative error 
$(\rho_k^{(num)}-\rho_k^{(approx)})/\rho_k^{(num)}$
    (for $k=10\, m$) as function of time (in unit of $1/m$).}
    \label{fig:error}
    \end{figure}

Secondly, we wish to show how accurate the approximation 
(\ref{xsolution}) is mode by mode.
In order to do this we numerically solve Eq. (\ref{pinneyrho})
and compare it to the approximation employed.
The agreement is very good up to times very close to the end of
inflation.
In Fig. \ref{fig:error} we show for the mode with $k=10 \, m$
the time evolution of the relative  error
$(\rho_k^{(num)}-\rho_k^{(approx)})/\rho_k^{(num)}$.
The larger $k$ is the better the agreement.   
We are therefore allowed to use the approximation for the inflaton 
dynamics almost til the end of inflation.

Let us now consider the correlator $\langle \varphi^2 \rangle$
(similar considerations are valid for other quantities bilinear in the field).
We note from the spectral analysis that
the integral over the modes can be split into two parts
    \be
    \langle \varphi^2 \rangle =
    \frac{\hbar}{(2 \pi)^3} \int d^3 {\bf k} \, |\varphi_{\bf k}|^2
    = \frac{\hbar}{2 (2 \pi)^3} \int d^3 {\bf k} \, x_k^2 = \frac{\hbar}
    {4 \pi^2} \left[ \int_0^{\ell} d k \, k^2 \, x_k^2 + \int_{\ell}^{\infty} d
    k \, k^2 \, x_k^2 \right]
    \,
    \label{fisquare}
    \ee
    
    Below the scale $\ell$ both the interval with the oscillations and the
    tilt around the scale invariant spectrum shown in Fig. 1 are included.
    The analytic treatment of the
    far infrared modes which contribute to the first integral would need an
    analytic approximation for the modes which
    at the beginning of inflation are outside the Hubble radius. This would
    amount to knowing the phase and the initial quantum states which precede
    the inflationary phase. Therefore we shall proceed by considering only the
    second integral and we shall neglect the first one in the far infrared.
    Even on neglecting the first integral which would require extra
    assumptions, the correct leading behaviour of the renormalized quantities
    is obtained \cite{explanation}.

    We conclude this section by noting that other methods to deal with
    infrared states are present in the literature. Infrared states were
    studied by Ford and Parker \cite{fordparker} for massless fields in
    Robertson-Walker space-times with a power-law expansion of a particular
    kind. By matching an earlier static
    space-time with a space-time with a scale factor which expands in time
    with a power law, they noted that an infrared finite state cannot evolve
    to an
    infrared divergent state \cite{fordparker}.
    The same scheme was also used by Vilenkin and Ford \cite{VF} for the
    problem of massless minimally coupled scalar fields in de Sitter
    space-time. An earlier radiation dominated phase was matched to the de
    Sitter metric. Through this matching the infrared tail becomes 
suppressed leading to an infrared finite state. 
As we shall see both the calculations performed by eliminating the 
infrared tail (i.e. working with the cut-off) 
and suppressing the infrared tail (through the Bogoliubov coefficients 
obtained by the matching prescription) lead to the same result to leading 
order. Indeed, their physical motivation is the same:
    inflation is not eternal, but starts at a finite time. However, the two
    methods treat the infrared tail in a different way: the agreement to
    leading order implies that the relevant contribution to the correlator
    $\langle \varphi^2 \rangle$ -
    and to the EMT - comes from intermediate modes, and not from the
    furthest infrared modes.

    \section{The energy-momentum tensor} \label{four}

    The classical energy-momentum tensor (henceforth EMT) of inflaton
    fluctuations is:
    \be
    T_{\mu\nu} = \partial_\mu \varphi \partial_\nu \varphi
    + g_{\mu\nu} \left[- \frac{1}{2} g^{\alpha\beta} \partial_\alpha
    \varphi \partial_\beta \varphi - \frac{m^2}{2} \varphi^2 \right]
    \label{emt}
    \ee
    and its operator form is simply obtained by promoting $\varphi$ to an
    operator as in Eq. (\ref{quantumFourier}).

    When averaged over the vacuum state annihilated by $b$:
    \be
    \hat{b}_{\bf k} | 0 \rangle = 0
    \ee
    the energy-momentum tensor assumes a perfect fluid form because of the
    symmetries of the RW background \cite{guven}:
    \be
    \langle T_{\mu \nu} \rangle = {\rm diag} (\epsilon, a^2 p \delta_{ij})
    \,,
    \ee
    where $\epsilon$ and $p$ are the energy density and the pressure density
    respectively.

    In the following we consider, according to the previous sections, $\nu=3/2$,
    employing the dimensional regularization \cite{birrell} to treat the UV behaviour.
    Therefore the integrands will be in $3$ dimensions and the integration measure
    analitically continued in $d$ dimension.

    The energy density is
    \begin{eqnarray}
    \epsilon = \langle T_{00} \rangle &=&
    \frac{\hbar}{2 (2 \pi)^d} \int_{|k|>\ell} d^d {\bf k} \left[
    |\dot\varphi_{\bf k}|^2
    + \frac{k^2}{a^2} |\varphi_{\bf k}|^2 + m^2 |\varphi_{\bf k}|^2
    \right] \nonumber \\
    &=& \frac{\hbar}{4 (2 \pi)^d}
    \int_{|k|>\ell}
    d^d {\bf
    k} \left[ {\dot x}_k^2 + \frac{1}{a^6 x_k^2} + \left( \frac{k^2}{a^2} +
    m^2 \right) x_k^2 \right] \,,
    \label{endensity}
    \end{eqnarray}
    and the pressure density $p$, related to the space-space component of the EMT, is:
    \be
    p = \frac{\langle T_{ii} \rangle}{a^2} = \frac{\hbar}{4 (2
    \pi)^d}
    \int_{|k|>\ell} d^d {\bf k} \left[ {\dot x}_k^2 + \frac{1}{a^6 x_k^2} +
    \left(
    \frac{2-d}{d} \frac{k^2}{a^2} - m^2 \right) x_k^2 \right] \,.
    \label{pressdensity}
    \ee

    On using Eqs. (\ref{xsolution}) and (\ref{prud1}) in Appendix B
    with $\alpha = d-1$, we may
    now compute the second part of the integral (\ref{fisquare}):
    \begin{eqnarray}
    \langle \varphi^2 \rangle &=&
    \frac{\hbar}{2 (2 \pi)^d} \int_{|k|>\ell} d^d {\bf k} \, x_k^2
    \label{newphiB} \\
    &=&\frac{\hbar}{16\pi^2}
    H^2 \left( 1-\frac{2}{3}\frac{m^2}{H^2}\right) \Bigl\{ 2-4 \ln 2-
    2 \left(\frac{\ell}{2 \pi^{1/2}} \right)^{d-3}
    \Gamma \left(\frac{1}{2}-\frac{d}{2}\right)   + \nonumber \\
    & &
    +{\cal O} \left (
    \frac{1}{a^2} \right )\Bigr\} +{\cal O} (d-3)
    \label{fisquare_paper}
    \end{eqnarray}

    Analogously, the energy and pressure densities in Eqs.
    (\ref{endensity},\ref{pressdensity}), with the help of the formulae
    in Appendix B, are:
    \begin{eqnarray}
    \epsilon=\langle T_{0 0} \rangle = \frac{\hbar}{16\pi^2} H^4 &\Bigl\{&
    4 \frac{m^2}{H^2}
    \left( \frac{1}{2 \pi^{1/2}} \right)^{d-3}
    \ell^{d-3} \Gamma(1-d) \nonumber \\
    & &+\frac{m^2}{H^2} \Bigl[ -1
    +\gamma -2 \ln 2 \Bigr] +{\cal O}
    \left( \frac{1}{a^2} \right) \Bigr\} +{\cal O} (d-3)
    \label{bare_energy}
    \end{eqnarray}
    \begin{eqnarray}
     p = \frac{\langle T_{i i} \rangle}{a^2} =
    \frac{\hbar}{16\pi^2} H^4 &\Bigl\{&
    - 4 \frac{m^2}{H^2}  \left (\frac{1}{2 \pi^{1/2}} \right
    )^{d-3}\ell^{d-3} \Gamma(1-d) \nonumber \\
    & & +\frac{m^2}{H^2} \Bigl[ +1 -\gamma +2 \ln 2 \Bigr] +{\cal O} \left (
    \frac{1}{a^2} \right )\Bigr\} +{\cal O} (d-3)
    \label{bare_pressure}
    \end{eqnarray}

    The poles given by the negative values of the argument of the $\Gamma$
    function in Eqs. (\ref{fisquare_paper},\ref{bare_energy},\ref{bare_pressure}) represent part
    of the ultraviolet infinities of field theory.

    \section{The adiabatic subtraction} \label{five}

    In order to remove the divergent quantities which appear in the integrated
    quantities as poles in the $\Gamma$ functions, we shall employ the method
    of {\em adiabatic subtraction} \cite{adiabatic}. Such a method consists in
    replacing $x_k$ with an expansion in powers of derivatives of the
    logarithm of the scale factor in Eqs. (\ref{endensity}-\ref{newphiB}).
    This expansion
    coincides with the adiabatic expansion introduced by Lewis in \cite{lewis}
    for a time dependent oscillator.

    Usually it is more convenient to formulate the adiabatic expansion by
    using the conformal time $\eta$ \cite{adiabatic} ($d \eta = d t/a $).
    We follow this procedure and write an expansion in derivatives with
    respect to the conformal time (denoted by $'$) for $x_k$. Then go back to
    the cosmic time and we insert the expansion in the expectation values we wish to
    compute.  Adiabatic expansion in cosmic time and
    conformal time lead to equivalent results, because of
    the explicit covariance under time reparametrization \cite{HMPM}.

    We rewrite Eq. (\ref{pinneyx}) in conformal time in the following
    way:
    \be
    (a x_k)'' + \Omega_k^2 \, (a x_k) =
    \frac{1}{(a x_k)^3}
    \label{conformal_eq}
    \ee
    where
    \be
    \Omega_k^2 = k^2 + m^2 a^2 - \frac{1}{6} a^2 R
    \label{conformal_freq}
    \ee
    and $R$ is the Ricci  curvature:
    \be
    R = 6 \frac{a ''}{a^3}   \,.
    \label{ricci_d}
    \ee

    The fourth order expansion for $\langle \varphi^2\rangle$, the energy and
    pressure densities are
    therefore (as before $\nu=3/2$, integrands in $3$ space dimensions and
    the measure is analytically continued in $d$ dimensions), using the expression in
    (\ref{fourth_d}) and the results of appendix A and B,

    \begin{eqnarray}
    \langle \varphi^2\rangle_{(4)} &=&\frac{\hbar}{16\pi^2}  H^2
    \Bigl\{ \left[ -2+\frac{4}{3}\frac{m^2}{H^2}\right]
    \left(\frac{a m}{2 \pi^{1/2}} \right)^{d-3} \Gamma
    \left(\frac{1}{2}-\frac{d}{2}\right) +\frac{2}{9}
    \frac{m^2}{H^2}-\frac{4}{3}  + \nonumber \\
    & &
    +\frac{1}{m^2}\left[\frac{7}{45}m^2+\frac{29}{15}H^2\right]+{\cal O} \left
    ( \frac{1}{a^3} \right )\Bigr\} +{\cal O} (d-3)
    \label{fisquare_appendix2}
    \end{eqnarray}

    \begin{eqnarray}
    \epsilon_{(4)}=\langle T_{0 0} \rangle_{(4)} = \frac{\hbar}{16\pi^2} H^4 &\Bigl\{&
    4 \frac{m^2}{H^2}  \left (\frac{1}{2 \pi^{1/2}}
    \right )^{d-3}a^{d-3}m^{d-3} \Gamma(1-d) \nonumber \\
    & & + \frac{119}{60}
    +\frac{m^2}{H^2} \left [ -\frac{33}{10}
    +\gamma \right ] +{\cal O} \left
    ( \frac{1}{a^3} \right )\Bigr\} +{\cal O} (d-3)
    \label{en_fourth}
    \end{eqnarray}
    \begin{eqnarray}
    p_{(4)}=\frac{\langle T_{i i}\rangle_{(4)}}{a^2} = \frac{\hbar}{16\pi^2}  H^4 &\Bigl\{&
    -4 \frac{m^2}{H^2}  \left (\frac{1}{2 \pi^{1/2}}
    \right )^{d-3}a^{d-3}m^{d-3} \Gamma(1-d) \nonumber \\
    & & - \frac{119}{60} +
    \frac{m^2}{H^2} \left [ +\frac{1309}{270}
    -\gamma \right ] +{\cal O} \left
    ( \frac{1}{a^3} \right )\Bigr\} +{\cal O} (d-3)
    \label{press_fourth}
    \end{eqnarray}


    \section{The conserved renormalized EMT}

    On subtracting the adiabatic part given in Section \ref{five} from the
    bare integrated contribution given in Section \ref{four}
    and taking the limit
    $d \rightarrow 3$ one obtains the finite renormalized expectation value
    for the correlator and for the energy-momentum tensor.

    The renormalized expectation value of $\langle \varphi^2 \rangle$ is
    therefore, neglecting terms of order $1/a^3$ and for $a> H/m$ \cite{explanation}, 
    \begin{eqnarray}
    \langle \varphi^2 \rangle_{REN} &=&   \langle \varphi^2\rangle-
    \langle \varphi^2 \rangle_{(4)}  \nonumber \\
    &=& \frac{\hbar}{16\pi^2}  H^2
    \Bigl\{ \left( 4-\frac{8}{3}\frac{m^2}{H^2}\right) \left( \ln a-
    \ln \frac{C H(t_0)}{m} \right) -\left( 1-\frac{2}{3}\frac{m^2}{H^2}\right)
    4 \ln 2- \nonumber \\
    & & -\frac{14}{9}\frac{m^2}{H^2}+\frac{10}{3}
    - \frac{1}{m^2}\left[\frac{7}{45}m^2+\frac{29}{15}H^2\right]
    + {\cal O} \left (\frac{1}{a^3} \right )\Bigr\}\; .
    \label{fisquare_ren}
    \end{eqnarray}

Considering (\ref{fisquare_ren}), we note that for late times it resembles 
more a massless, than a massive, field in de Sitter space-time.
This is a consequence of the scale invariant spectrum 
(\ref{parameters2}) of inflaton fluctations (the same spectrum occurs for 
massless minimally coupled fields in de Sitter space-time). The 
leading behaviour for $\langle \varphi^2 \rangle_{REN}$ agrees for 
late times with 
the de Sitter result \cite{AF,VF,cutoff} 
when $\dot H=0$ and $a(t) = a_0 e^{H_{\rm DS} t}$:   
\be
\langle \varphi^2 \rangle_{\rm REN}^{\rm DS} \sim
\frac{\hbar}{4 \pi^2} H_{\rm DS}^3 t
\,. \label{afleading}
\ee
At earlier times, $\langle \varphi^2 \rangle_{REN}$ is dominated by a 
contribution ${\cal O} (H^4/m^2)$, but then the time dependent piece 
${\cal O}(\ln a)$ takes over. We warn the reader about the massless limit 
taken at face value of Eq. (\ref{fisquare_ren}). In the massless limit, as 
discussed in the appendix B, one has to use a different analytic 
continuation for the adiabatic part, related to the expression in 
(\ref{intadiamassless}) and the result for $\langle \varphi^2 
\rangle_{REN}$ will be finite, different from Eq. 
(\ref{fisquare_ren}), but with the same leading contribution (\ref{afleading}).
However, this massless limit is just of academic 
interest, since for $m=0$ inflation would not happen. 

    Even if both inflaton fluctuations for a massive inflaton and a
    massless minimally coupled scalar fields in de Sitter share the same
    scale invariant spectrum, the
    energy and pressure carried by fluctuations for these two cases are very
    different. For the latter case,
    a linear growth in time is present only for the correlator; the EMT does
    not contain the correlator, but only bilinear quantities less infrared
    than $\varphi^2$ (for a nice explanation of this difference see
    \cite{HMPM}). For the case of inflaton fluctuations, the correlator
    appears directly in the EMT because of the nonvanishing mass. The kinetic
    and gradient terms should be smaller than the potential term, as for
    the massless minimally coupled case. Therefore for the case of inflation
    driven by a massive inflaton, the EMT of inflaton fluctuations should grow
    in time. This is what we shall show in the following.

    On subtracting Eq. (\ref{en_fourth}) from Eq. (\ref{bare_energy}),
    the renormalized energy density $\epsilon_{\rm REN}$ is:
    \begin{eqnarray}
    \epsilon_{\rm REN} &=& \langle T_{00} \rangle_{\rm REN} =
    \langle T_{00} \rangle - \langle T_{00} \rangle_{(4)} = \nonumber \\
    &=&\frac{\hbar}{16\pi^2} H^4 \Bigl\{
    -2 \frac{m^2}{H^2} \left [ \ln \frac{C H(t_0)}{m}-\ln a(t)\right ]-
    \frac{119}{60}
    +\frac{m^2}{H^2} \Bigl[\frac{23}{10}
    - 2 \ln 2 \Bigr] +{\cal O} \left
    ( \frac{1}{a^2} \right )\Bigr\}\,.
    \label{energy_ren}
    \end{eqnarray}
    Similarly, by subtracting Eq. (\ref{press_fourth}) from Eq.
    (\ref{bare_pressure}), the renormalized pressure density $p_{\rm REN}$ is:
    \begin{eqnarray}
    p_{\rm REN} &=& \frac{\langle T_{i i} \rangle_{\rm REN}}{a^2} =
    \frac{ \langle T_{ii} \rangle - \langle T_{ii} \rangle_{(4)} }{a^2}
    \nonumber \\
    &=&\frac{\hbar}{16\pi^2} H^4 \Bigl\{
    2 \frac{m^2}{H^2} \left [ \ln \frac{C H(t_0)}{m}-\ln a(t)\right ]+
    \frac{119}{60}
    +\frac{m^2}{H^2} \Bigl[ -\frac{1039}{270}
    + 2 \ln 2 \Bigr] + {\cal O} \left
    ( \frac{1}{a^2} \right )\Bigr\}
    \label{pressure_ren}
    \end{eqnarray}

    We note that this result does not agree for $\dot H=0$ (and therefore
    for $m=0$ because of Eq. (\ref{hdot})) and $\xi=0$
    with the de Sitter result \cite{birrell,guven} obtained with the
    Bunch-Davies vacuum:
    \begin{eqnarray}
    T_{\mu \nu \, {\rm REN}}^{{\rm DS}\,{\rm BD}} &=& - \frac{g_{\mu
    \nu}}{64
    \pi^2} \left[ m^2
    \left[ m^2 + (\xi - \frac{1}{6}) R \right]
    \left[ \psi(\frac{3}{2} + \nu) + \psi(\frac{3}{2} - \nu) -
    \ln \frac{12 m^2}{R} \right] - \right. \nonumber \\
    && \left. - m^2 (\xi - \frac{1}{6}) R
    - \frac{1}{18} m^2 R - \frac{1}{2} (\xi - \frac{1}{6})^2 R^2
    + \frac{1}{2160} R^2 \right]
    \label{desitter}
    \end{eqnarray}
    where $R$ is the curvature in de Sitter ($R = 12 H^2_{\rm DS}$ with
    $H_{\rm DS}$ as the Hubble parameter in de Sitter space-time)
    \cite{largem,senzacutoff}.
    Indeed, the limit of Eq. (\ref{desitter}) for vanishing mass and coupling
    is finite and is \cite{AF}:
    \be
    T_{\mu \nu \, {\rm REN}}^{{\rm DS}\, {\rm BD}} = -
    \frac{61 H^4_{\rm DS}}{960 \pi^2} g_{\mu \nu}
    \label{desitter_BD}
    \ee
    Instead, the result (\ref{energy_ren},\ref{pressure_ren}) agrees with
    the Allen-Folacci \cite{AF} result for $m=0$:
    \be
    T_{\mu \nu \, {\rm REN}}^{{\rm DS}\, {\rm AF}} =
    \frac{119 H^4_{\rm DS}}{960 \pi^2} g_{\mu \nu}
    \label{desitter_AF}
    \ee

    However, the renormalized EMT of inflaton fluctuations in
    Eqs. (\ref{energy_ren},\ref{pressure_ren}) grows in time, as the logarithm
    of the scale factor. This feature is due
    both to the fact that the inflaton is massive and to the infrared state in
    which its fluctuations are generated. For a test field in de Sitter
    space-time a linear growth in time of the EMT is possible,
    only for $m^2 + \xi R = 0$, with $m$ and $\xi$ both different from zero.

    The renormalized EMT in de Sitter space-time in the
    Bunch-Davies vacuum (\ref{desitter}) and in the Allen-Folacci vacuum
    corresponds to a perfect fluid with an equation of state $w = p/\epsilon
    =-1$, which is identical to the background driven by a cosmological
    constant. The conservation of the renormalized
    EMT (\ref{desitter},\ref{desitter_AF})
    is direct consequence of its symmetries:
    \be
    T_{\mu \nu \, {\rm REN}}^{\rm DS} \propto g_{\mu \nu} \rightarrow
    \nabla^\mu T_{\mu \nu \, {\rm REN}}^{\rm DS} = 0
    \ee
    since $\nabla^\mu g_{\mu \nu} = 0$.

    The renormalized EMT in an inflationary stage driven by a
    massive inflaton corresponds to a perfect fluid, {\em but}
    with an equation of state which differs from $-1$ by terms
    ${\cal O} (m^2/H^2)$, as one can see from Eqs.
    (\ref{energy_ren},\ref{pressure_ren}).

    The derivation of the conservation of the renormalized EMT in chaotic
    inflation is also straightforward. The renormalized EMT is conserved
    consistently with the approximation used, i.e. to the order ${\cal O}
    (m^2/H^2)$. The conservation can be easily
    checked mode by mode, i.e. by considering the covariant
    derivative inside the integrals in $\bf k$ in the difference
    between the bare value and the fourth order adiabatic value and
     using the equations of motion for the field modes (\ref{scalar1}).
     The conservation of the final renormalized value EMT can also
    be checked by inserting the expressions given by Eqs.
    (\ref{energy_ren}) and (\ref{pressure_ren}) in
    \be
    \frac{d \epsilon_{\rm REN}}{d t} +
    3 H (\epsilon_{\rm REN} + p_{\rm REN} ) = 0 \,.
    \label{cons_ren}
    \ee
    and retaining only the terms up to and including ${\cal O} (m^2/H^2)$.

    \section{Back-reaction on the geometry}

    We now discuss the back-reaction of the
    amplified $\varphi$ fluctuations on the geometry.
   
    We consider the back-reaction equations perturbatively:
    we evaluate the higher order
    geometrical terms \cite{birrell} generated by renormalization as their
    background value and we do not use them to generate higher order
    differential equations.
    We note that, in accord with the approximations used, many higher order
    derivative terms are implicitly absent
    in Eqs. (\ref{energy_ren},\ref{pressure_ren}),
    since they would be of higher order in
    powers of ${\dot H}/H^2$ and because $\ddot H \simeq 0$.
    We estimate the back-reaction effects without changing
    the structure of the left hand side of Einstein equations.
    Hence the back-reaction equations we consider are:
    \begin{eqnarray}
    H^2 &=& \frac{8 \pi G}{3} \left[ \frac{\dot \phi^2}{2} + \frac{m^2}{2}
    \phi^2
    + \epsilon_{\rm REN} \right] \label{hubble_back} \\
    \dot H &=& - 4 \pi G ( \dot \phi^2 + \epsilon_{\rm REN} + p_{\rm REN})
    \label{hdot_back} \,.
    \end{eqnarray}

    The main point is that the energy and pressure of inflaton fluctuations
    grows in time as the logarithm of the scale factor, while the 
    Hubble parameter driven by the background energy density decreases linearly 
    in time. However, the
    approximation we use, i.e. Eqs. (\ref{hdot},\ref{evolution}), is valid for
    a certain time interval, $\Delta t$ given by:
    
    \be
    \Delta t \sim \frac{H_0}{m^2}
    \label{interval}
    \ee
    
    Therefore the term which grows in the renormalized EMT will saturate at
    the value:
    \be
    \epsilon_{\rm REN} (\Delta t) \sim - p_{\rm REN} (\Delta t)
    \sim \frac{\hbar}{8 \pi^2} H^2 H_0^2 \,.
    \label{maximum}
    \ee

\begin{figure}
    \leavevmode\epsfysize=2.5in \epsfbox{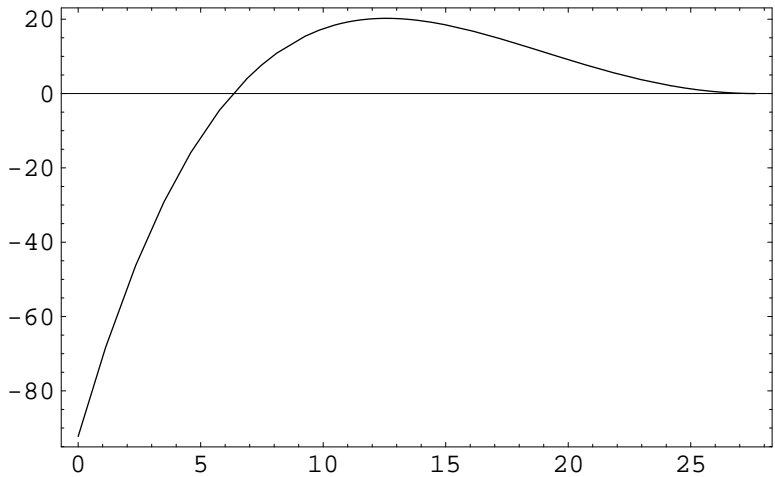}
    \caption{Time evolution of the renormalized energy,
    where the cosmological time is in units of $1/m$.}
    \label{fig:eren}
    \end{figure}

\begin{figure}
    \leavevmode\epsfysize=2.5in \epsfbox{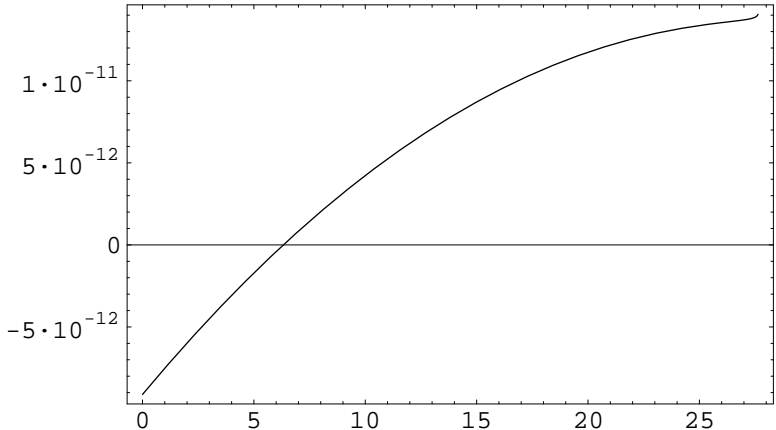}
    \caption{Time evolution of the magnitude of the back-reaction
    $8\pi\,G\,\epsilon_{REN}/(3H^2|_{\epsilon_{REN}=0})$,
    where the cosmological time is in unit of $1/m$.}
    \label{fig:ratio_br}
    \end{figure}

    The term (\ref{maximum}) is larger than the Allen-Folacci value
    $\sim H^4$ in Eq. (\ref{desitter_AF}) and has the opposite sign.
    Therefore,
    the energy density of fluctuations starts with a negative value and
    changes sign to a positive value when the logarithm takes over. This
    behaviour is shown in Figs. \ref{fig:eren} and \ref{fig:ratio_br}.
    
    Such a value leads to a contribution to the Einstein
    equations of the order $H^2 H^2_0/M_{\rm pl}^2$. The importance of
    back-reaction is therefore related to the ratio $H^2_0/M_{\rm pl}^2$. If
    inflation starts at a Planckian energy density then back-reaction 
during inflation cannot be neglected.

    A sligthly different conclusion can be reached on considering the
    variation in time
    of the Hubble parameter, i.e. Eq. (\ref{hdot_back}). The important point
    to note is that the leading contribution in $\epsilon_{\rm REN}, p_{\rm
    REN}$, i.e. the terms $\sim H^4$ and $\sim m^2 H^2 \log a$, do not
    contribute to $\dot H$ since these terms have an
    equation of state $p_{\rm REN}/\epsilon_{\rm REN}=-1$. The contribution
    of the inflaton fluctuations to $\dot H$ is therefore of order
    $m^2 H^2/M_{\rm pl}^2$, which is suppressed with respect to the classical
    value by the factor $H^2/M_{\rm pl}^2$.

    \section{Comparison with the slow-rollover calculation}

    We now compare the approximation which lead us to Eq.
    (\ref{solution}) with the slow-rollover technique
    introduced
    by Stewart and Lyth \cite{SL}. The latter technique was developed directly
    for scalar and tensor metric perturbations \cite{SL}, and not for field
    perturbations in rigid space-times, as treated here. However, the equation
    for gravitational waves differ from Eq. (\ref{scalar1}) only by the
    presence of the mass term $m$. Therefore, as a first check we note
    from Eq. (\ref{parameters2}) that
    \be
    \nu^2_{\rm GW} = \frac{9}{4} - 3 \frac{\dot H}{H^2}
    \quad \nu_{\rm GW} \simeq \frac{3}{2} - \frac{\dot H}{H^2}
    \label{gw}
    \ee
    where the second relation holds when ${\dot H}/H^2$ is small. The value
    $\nu_{\rm GW}$ in Eq. (\ref{gw}) coincides with the index $\mu$ of Eq.
    (41) of \cite{SL}. On the other hand, if one applies the Stewart-Lyth
    procedure to Eq. (\ref{scalar1}) one would obtain an index for Hankel
    functions which is precisely $\nu$ in Eq. (\ref{parameters2}).

    A natural question is to ask whether inflaton fluctuations are
    generated in infrared states also for other models of chaotic inflation.
    Analytic approximations, such as the one presented in
    Section 2, are very difficult to obtain. However,
    since our
    calculation agrees with the slow-rollover result \cite{SL} for 
inflaton fluctuations in rigid space-times and gravitational waves in the 
case of a massive inflaton, one can use
    the latter technique to estimate the spectral index of fluctuations in a
    generic chaotic model with potential $V(\phi) = \lambda \phi^n/n$ (here
    $\lambda$ has the dimensions of a mass elevated to the power $4 - n$).
    We follow Stewart and Lyth and we study the equation:
    \be
    (a \varphi_{\bf k})'' + \left( k^2 + m^2_{\rm eff} a^2 - \frac{a''}{a}
    \right) (a \varphi_{\bf k}) = 0 \,,
    \ee
    where $m^2_{\rm eff} = V,_{\phi \phi} = \lambda (n-1) \phi^{n-2}$.
    On using $a(\eta) = - 1/[ H \eta (1 - \epsilon)]$, after some algebra
    one obtains the following spectral index for the constant mode:
    \be
    \nu = \frac{1}{2} + \frac{1 - 2\epsilon + 2 \epsilon/n}{1 - \epsilon}
    \label{general_nu}
    \ee
    where the slow-rollover parameter $\epsilon$ is taken as constant and
    is defined as:
    \be
    \epsilon = - \frac{\dot{H}}{H^2} = \frac{n^2}{16 \pi G \phi^2}
    \ee
    The slow-rollover approximation is better for large values of the inflaton.
    The result (\ref{general_nu})
    agrees for $n=2$ with the quadratic case. However, for $n > 2$ the
    spectral index
    $\nu$ is smaller than the critical value $3/2$, leading to inflaton
    fluctuations which are infrared finite.

    Although the approximation presented here gives the same result as the
    slow-rollover technique to the lowest order \cite{SL} for gravitational
    waves and inflaton fluctuations in rigid space-times for a massive 
chaotic model, it will not be so when one includes metric perturbations
    \cite{new}.

    \section{Moduli Production}

    Let us now discuss the quantum production of a light scalar field $\chi$
    in this model of massive chaotic inflation. By the term light, we mean 
    a scalar field
    with a mass $M$, which is smaller than (or equal to) the inflaton one, $m$.
    We assume a vanishing homogeneous component for $\chi$. 
    Therefore, for $\chi$ the approximation of a rigid space-time is correct.

    With $M < m$, we see from Eqs. (\ref{hdot},\ref{parameters2}) that
    the index $\nu_\chi$ for the Hankel functions, which are involved in
    the solutions for the $\chi$ modes, is larger than $3/2$. The leading
    contribution for
    the renormalized value of $\langle \chi^2 \rangle$ is:
    \be
    \langle \chi^2 \rangle_{\rm REN} \sim \frac{1}{16 \pi^2} \beta H^2 \;
    \frac{a^{2 \nu_\chi - 3}-1}{2 \nu_\chi - 3}
    \label{chisquare_ren}
    \ee
    where $\beta$ is a numerical coefficient.
This formula generalizes Eq.
    (\ref{fisquare_ren}) to the case of a mass $M < m$. The case $M=m$
    represents a limiting case, for which a logarithm appears. Analogously, in
    de Sitter space-time, a limiting case is also present for $\nu = 3/2$
    \cite{HMPM}, and a logarithm of the scale factor appears instead of a
    power.

    For a massive $\chi$ the following relation holds:
    \be
    \nu_{\rm GW} > \nu_\chi > \nu = \frac{3}{2} \,
    \ee
    where $\nu_{\rm GW}$ is defined in Eq. (\ref{gw})
    and, owing to the smallness of the ratio $m^2/H^2$, $\nu_\chi$
    is very close to $3/2$. However, the growth in time of $\langle \chi^2
    \rangle_{\rm REN}$ is more rapid than the growth of
    $\langle \varphi^2 \rangle_{\rm REN}$.

This fact can be checked numerically: for example,
on taking the initial condition for inflation already used in the previous
sections for the other numerical checks, one finds for $M=0$, 
$\nu_\chi \simeq 1.5043$ and for $M=0.5m$,  $\nu_\chi \simeq 1.5032$.
Such a value is time independent after a very short transient phase
needed for the modes to freeze and corresponds to the one given in
(\ref{gw}) for $H$ computed
for a time very close to the beginning of the inflation. We can therefore
substitute $H(t \simeq 0) \simeq H_0$ in place of $H(t)$ in $\nu$.
We have checked this relation for different initial $H_0$.
This fact is in agreement with the fact that the approximate solution,
which we employed for the inflaton field, can be used for other fields
with different masses, but only at the beginning of the inflation period.
Thus we can control the spectral behavoiur of the moduli field, but
not its normalization.
  
To study the behaviour in (\ref{chisquare_ren})
it is convenient to rewrite the evolution of the
scale factor $a$ and its exponent $(2\nu_\chi-3)$ 
as
\be
a(t)=\exp{\left(\frac{3}{2}\frac{H_0^2-H^2}{m^2}\right)} \quad , \quad 
2\nu_\chi-3 \simeq \frac{2}{3}\frac{m^2-M^2}{H_0^2}\; .
\ee

Therefore the main result (let us only write the dominant contribution) is that
for $M < m$ and at the end of inflation
    \be
    \langle \chi^2 \rangle_{\rm REN}
\sim H^2 H_0^2
\left\{ \frac{\exp{\left[\frac{H_0^2-H^2}{H_0^2} 
\left(1-\frac{M^2}{m^2}\right)
\right]}-1}{m^2-M^2} \right\}
> \langle \varphi^2 \rangle_{\rm
    REN} \sim H^2 \frac{H_0^2}{m^2}
    \label{result1}
    \ee
and
for the renormalized EMT associated with the $\chi$ field
\be
 \epsilon_{\rm REN}^{(\chi)} \sim \frac{3}{16\pi^2}
\frac{M^2}{m^2} H^2 H_0^2 
\left\{ \frac{\exp{\left[\frac{ H_0^2-H^2}{H_0^2} 
\left(1-\frac{M^2}{m^2}\right)
\right]}-1}{1-\frac{M^2}{m^2}} \right\} \; .
\label{result2}
\ee 
    The results in Eqs. (\ref{maximum},\ref{result1}, \ref{result2})
are very interesting.
    They show that the production of a scalar field $\chi$
    with mass $M$ smaller or equal to the inflaton mass $m$ depends on the
    duration of inflation and is larger than the usual extrapolation of the
    de Sitter result ($\langle \chi^2 \rangle \sim H^4/m^2$ and $\epsilon_\chi
    \sim (M^2/m^2) H^4\;$).
This result implies that the quantum production of light 
scalar fields in
    chaotic inflation with a mass term is
    even greater than expected on extrapolating the de Sitter result, and
    depends on the duration of inflation, as is also stated in 
\cite{kuztka,moduli}. Again, if inflation starts at a Planckian energy 
density, the back-reaction of light scalar fields cannot be neglected during 
inflation.

We note that the factor in curly brackets in (\ref{result1},
\ref{result2}) is larger than the logarithmic term one has for the $\nu=3/2$
case, by a factor of $2$.

We see that on computing the exact numerical solution of the Pinney equation
for the modes of the inflaton with mass $m$ and for a moduli
field with $M=0.5m$, one obtains for the same momenta
amplitudes more than one order of magnitude larger for the latter at the end of
inflation, again in agreement with \cite{kuztka}.
We therefore expect an enhancement of 2-3 orders of magnitude
for its backreaction with respect to the inflaton case.
  
    \section{Discussion and Conclusions}

    We have computed the renormalized conserved EMT of the inflaton
    fluctuations $\varphi (t, {\bf x})$ in rigid space-times during the
    inflationary stage driven by a mass term.
    The method of dimensional regularization has been applied by using an
    analytic approximation valid during the slow-rollover regime.
    All the results agree with the Allen-Folacci results for
    $T_{\mu\nu}$ of a test field in de Sitter
    space-time \cite{AF}, in the limit for which the Hubble parameter is
    constant (which is also the massless limit because of Eq. (\ref{hdot})).

    We find that the EMT of inflaton fluctuations grows in time.
    The reason for this behaviour is that chaotic
    inflation driven by a massive scalar field produces a scale 
invariant spectrum of fluctuations even if the field is massive.
    This effect is due to the decrease of the Hubble parameter during
    the slow-rollover regime.

    In de Sitter space-time, the renormalized EMT of a
    quantum field grows linearly in time {\em only} if
    $m^2 + \xi R_{\rm DS} = 0$ with $m$
    and $\xi$ different from zero \cite{HMPM}.
    A massless minimally coupled scalar field in de Sitter
    space-time, characterized by a scale invariant spectrum of fluctuations,
    leads to a correlator which grows in time.
    However, only bilinear quantities less infrared than the correlator
    appear in the EMT, and therefore the expectation value of the EMT of a
    massless minimally coupled scalar field is constant in time \cite{HMPM}.
    In massive chaotic inflation,
    inflaton fluctuations are generated with a scale invariant spectrum.
    Since the correlator appears
    directly in the EMT because of the nonvanishing mass, then the 
renormalized EMT grows in time just as the
    correlator does.

    We find that the growth of the EMT of inflaton fluctuations during
    slow-rollover leads to a positive energy density which reaches a maximum
    value ${\cal O} (H^2 H^2_0)$, where $H_0$ is the Hubble radius at the
    beginning of inflation. This value exceeds the usual value
    ${\cal O} (H^4)$, which is of the same order of magnitude as the conformal
    anomaly. These values also show that back-reaction effects cannot be
    neglected if inflation starts at Planckian energies, i.e. at $H_0 \sim
    M_{\rm pl}$. If inflation started at Planckian energies, although the
    contribution of the terms ${\cal O} (H^2 H^2_0)$ and of the conformal
    anomaly would be of the same order of magnitude, we think that the two
    contributions could be different because of the different signs and of the
    different behaviours in time.

In this model of chaotic inflation, we have also analyzed the geometric 
production of an additional field $\chi$ with mass $M$ smaller than the 
inflaton mass $m$. Of course, on considering the normalization,
we have found that $\epsilon_\chi > 
\epsilon_\varphi \sim H^2 H_0^2$. This result implies that the quantum 
production of light fields depends on the duration of inflation and it is 
greater than expected on extrapolating the de Sitter result (in de 
Sitter $\epsilon_\chi \sim H^4$). As in the case of inflaton fluctuations, 
the energy density of light scalar fields could be comparable to the 
background one at the end of 
inflation, if inflation started at Planckian energy densities. Also in this 
case, the back-reaction of $\chi$ fluctuations does not appear to be
negligible during inflation.

    One may then ask whether this behaviour of
    the back-reaction due to the fluctuations in rigid space-time is
    common to other inflationary chaotic models. Analytic approximations, such as
    the one presented in Section 2, are very difficult to obtain. However,
    since our
    calculation agrees with the slow-rollover result \cite{SL} for the
    massive case, we have used the latter technique to estimate the spectrum
    of inflaton fluctuations in rigid space-time for a generic inflaton potential
    $V(\phi) = \lambda \phi^n/n$. We have found that for $n > 2$ the inflaton
    fluctuations are generated in an infrared finite state, leading to a
    back-reation which does not increase in time. However, we think that we
    must address the problem while including metric perturbations, in
    order to fully understand this issue. It is known that chaotic inflationary
    models predict a spectrum of curvature perturbations which is red tilted
    \cite{SL} - i.e. with a spectrum more infrared than the scale invariant
    one -, a result which does not hold for field perturbations in rigid
    space-time, as we have shown. Since infrared states could lead to
    a correlator which grows in time, the possibility exists that
    a back-reaction growing in time is common to all the chaotic inflationary
    models once metric perturbations are included.

    Other important issues are whether an eventual self-consistent scheme to
    include the back-reaction would prevent the development of infrared
    states.
    The effect of the self-consistent inclusion of back-reaction effects on
    the spectrum of fluctuations during inflation is, to our knowledge, an
    issue still to be fully explored. It would be interesting also to
    investigate the effect of the inclusion of the higher order terms
    \cite{simon} in the back-reaction equations
    (\ref{hubble_back},\ref{hdot_back}). Obviously, the Starobinsky model
    \cite{rsquare} is a surprising example of the importance of higher order
    terms.

    \vspace{0.5cm}
    \centerline{\bf Acknowledgments}
    \vspace{0.2cm}

    We would like to thank Raul Abramo, Robert Brandenberger, Sergei 
Khlebnikov and Igor Tkachev for
    discussions and comments on the manuscript. One of us (F. F.) would like
    to thank Salman Habib and Katrin Heitmann for many important discussions
    on renormalization in curved space-times and for warm hospitality at Los
    Alamos Laboratories, where part of this work was written.

    \section{Appendix A: The adiabatic fourth order expansion}

    From Eqs. (\ref{conformal_eq},\ref{conformal_freq},\ref{ricci_d}) one
    obtains the expansion for $x_k$ up to the fourth adiabatic order:
    \be
    x_{k}^{(4)}=\frac{1}{a}\frac{1}{\Omega_k^{1/2}}
    \left( 1-\frac{1}{4} \epsilon_2+\frac{5}{32}\epsilon_2^2-
    \frac{1}{4}\epsilon_4 \right)
    \label{fourth_1}
    \ee
    where $\Omega_k$ is defined in Eq. (\ref{conformal_freq}) and
    $\epsilon_2 \,, \epsilon_4$ are given by:
    \begin{eqnarray}
    \epsilon_2&=&-\frac{1}{2}\frac{\Omega_k^{''}}{\Omega_k^3}+\frac{3}{4}
    \frac{\Omega_k^{'2}}{\Omega_k^4} \nonumber \\
    \epsilon_4 &=& \frac{1}{4}\frac{\Omega_k^{'}}{\Omega_k^3}\epsilon_2^{'}-
    \frac{1}{4}\frac{1}{\Omega_k^2}\epsilon_2^{''}
    \end{eqnarray}

    The solution in Eq. (\ref{fourth_1}) must be expanded again since the
    Ricci curvature is of adiabatic order 2. Therefore $x_k^{(4)}$ is:
     \begin{eqnarray}
x_{k (4)} &=& \frac{1}{c^{1/2}}\frac{1}{\Sigma^{1/2}} \Bigl\{
1+\frac{1}{4}c\frac{R}{6}\frac{1}{\Sigma_k^2}+\frac{5}{32}c^2\frac{R^2}{36}
 \frac{1}{\Sigma_k^4}+ \nonumber \\
 & & +\frac{1}{16}\frac{1}{\Sigma_k^4}\left[c^{''} \left(m^2-\frac{R}{6}
 \right)-2 c^{'} \frac{R^{'}}{6}-c\frac{R^{''}}{6}\right]- \nonumber \\
 & &  -\frac{5}{64}\frac{1}{\Sigma_k^6}\left[ c^{'2}m^4-2c^{'2}m^2
 \frac{R}{6}-2 c^{'}m^2c\frac{R^{'}}{6}\right]+ \nonumber \\
 & & +\frac{9}{64}\frac{1}{\Sigma_k^6} c \frac{R}{6} c^{''}m^2-\frac{65}{256}  \frac{1}{\Sigma_k^8}
 c \frac{R}{6} c^{'2}m^4+\frac{5}{32}\epsilon_{2*}^2-
\frac{1}{4}\epsilon_{4*}   \Bigr\}
\end{eqnarray}

    where $c=a^2$ and
    \begin{eqnarray}
    \Sigma_k&=&(k^2 + a^2 m^2)^{1/2} \nonumber \\
    \epsilon_{2*}&=&-\frac{1}{2}\frac{\Sigma_k^{''}}{\Sigma_k^3}+\frac{3}{4}
    \frac{\Sigma_k^{'2}}{\Sigma_k^4} \nonumber \\
    \epsilon_{4*} &=&
    \frac{1}{4}\frac{\Sigma_k^{'}}{\Sigma_k^3}\epsilon_2^{'}-
    \frac{1}{4}\frac{1}{\Sigma_k^2}\epsilon_2^{''}
    \end{eqnarray}

    \section{Appendix B: Dimensional Regularization with cut-off}

    We start with  $\varphi^2$ as an example. According to the discussion
    following Eq. (\ref{fisquare}), we neglect the infrared piece of the integral
    since it gives a small finite part. Therefore the relevant integral
    \begin{eqnarray}
    \langle \varphi^2\rangle &=& \frac{1}{(2 \pi)^3}\frac{\hbar}{2} \frac{2
    \pi^{3/2}}{\Gamma
    (3/2)} \int_\ell^{+\infty} dk \, k^{2}
    x_k^2   \nonumber \\
    &=&
    \frac{\hbar}{(2 \pi)^3}\frac{\lambda}{a^3}
    \frac{\pi}{4 H} \frac{2 \pi^{3/2}}{\Gamma
    (3/2)} \int_\ell^{+\infty} dk \, k^{2} \left[J_\nu^2
    \left(\frac{\lambda k}{a H}\right)+
    N_\nu^2\left(\frac{\lambda k}{a H} \right) \right]
    \end{eqnarray}
    on extending it to $d$-dimensions (integrands in $3$ space dimensions
    and analytic continuation of the measure to $d$ dimensions) :

    \begin{eqnarray}
    \langle \varphi^2\rangle &=&
    \frac{\hbar}{(2 \pi)^d}\frac{\lambda}{a^3} \frac{\pi}{4 H}
    \frac{2 \pi^{d/2}}{\Gamma
    (d/2)} \int_\ell^{+\infty} dk \, k^{d-1}
    \left[J_\nu^2\left(\frac{\lambda k}{a H}\right)+
    N_\nu^2\left(\frac{\lambda k}{a H} \right) \right] \,.
    \end{eqnarray}
    The two-point function can be computed by using the following integral
    \cite{prudnikov}:
    \begin{eqnarray}
    & & \int_\frac{\lambda \ell}{a H}^{+\infty}dx
    x^{\alpha}[J_\nu^2(x)+N_\nu^2(x)]
    = \nonumber \\
    & &  \frac{1}{\pi^2}\Bigl\{\pi^{1/2}\left[
    \cos \left(\frac{\pi}{2}(\alpha+1-2\nu)\right)
    \frac{\Gamma ((\alpha+1)/2) \Gamma ((\alpha+1)/2-\nu)}{\Gamma
    (1+\alpha/2)}
    +\pi \frac{\Gamma (-\alpha/2)}{\Gamma
    ((1-\alpha)/2) \Gamma ((1-\alpha)/2+\nu)} \right]  \nonumber \\
    & & \Gamma (\frac{\alpha+1}{2}+\nu)+\frac{1}{(\alpha+1)\nu}\Bigl[ 2 \pi
    \left( \frac{\lambda \ell}{a H}\right)^d \cot (\pi \nu)
    _2F_3 \Bigl(\frac{1}{2},\frac{\alpha+1}{2};\frac{\alpha+3}{2},1-\nu,
    \nonumber \\
    & & 1+\nu;-\left(
    \frac{\lambda \ell}{a H}\right)^2\Bigr)\Bigr]+\frac{1}{2 \nu- \alpha -1}
    \Bigr[ 4^{\nu}\left(
    \frac{\lambda \ell}{a H}\right)^{\alpha+1-2 \nu} \Gamma (\nu)^2 \nonumber \\
    & & _2F_3 \left( \frac{1}{2}-\nu,\frac{\alpha+1}{2}-\nu;1-2\nu,1-\nu,
    \frac{\alpha+3}{2}-\nu;
    -\left(\frac{\lambda \ell}{a H}\right)^2\right)\Bigr]- \nonumber \\
    & &  -\frac{1}{(\alpha+1+2\nu)\Gamma (1+\nu)^2}\Bigl[ 4^{-\nu}  \left(
    \frac{\lambda \ell}{a H}\right)^{\alpha+1+2 \nu}
    (2 \pi^2+\cos(2 \nu \pi) \Gamma (-\nu)^2 \Gamma (1+\nu)^2 )
    \nonumber \\
    & &  _2F_3 \left(\frac{1}{2}+\nu,\frac{\alpha+1}{2}+\nu;1+\nu,
    \frac{\alpha+3}{2}+\nu,1+2 \nu;
    -\left(\frac{\lambda \ell}{a H}\right)^2\right) \Bigr] \Bigr\}
    \label{prud1}
    \end{eqnarray}

    With $\nu=3/2$ and $\alpha = d -1$, and on using:
    \begin{eqnarray}
    _2F_3 \left( b,c;d,e,f;-\left(\frac{\lambda \ell}{a H}\right)^2 \right)
    &=& 1+ {\cal O} \left( \frac{1}{a(t)^2}
    \right)
    \label{fact}
    \end{eqnarray}
    the expectation value for $\langle \varphi^2 \rangle$ is:
    \begin{eqnarray}
    \langle \varphi^2\rangle &=&\frac{\hbar}{16\pi^2}
    H^2 \left( 1-\frac{2}{3}\frac{m^2}{H^2}\right) \Bigl\{ 2-4 \ln 2-
    2 \left(\frac{\ell}{2 \pi^{1/2}} \right)^{d-3}
    \Gamma \left(\frac{1}{2}-\frac{d}{2}\right)   + \nonumber \\
    & &
    +{\cal O} \left (
    \frac{1}{a^2} \right )\Bigr\} +{\cal O} (d-3)
    \end{eqnarray}

    Similarly, the integral used for the fourth order adiabatic
    quantities in $d$-dimensions is (integrands in $3$ space dimensions and
    analytic continuation of the measure to $d$ dimensions):
    \begin{eqnarray}
    & & \frac{2 \pi^{\alpha/2+1/2}}{\Gamma ((\alpha+1)/2)}
    \int_\ell^{+\infty}
    dk k^{\alpha}\frac{1}{(k^2+a^2m^2)^{n/2}} =
    \nonumber \\ & & =\frac{2 \pi^{(\alpha+1)/2}}{\Gamma ((\alpha+1)/2)}
    \int_0^{+\infty} dk k^{\alpha}
    \frac{1}{(k^2+a^2m^2)^{n/2}}-
    \frac{2 \pi^{\alpha/2+1/2}}{\Gamma ((\alpha+1)/2)}
    \int_0^{\ell} dk k^{\alpha}\frac{1}{(k^2+a^2m^2)^{n/2}} \nonumber \\
    & &  =\pi^{\alpha/2+1/2} (a^2 m^2)^{\alpha/2+1/2-n/2}\frac{\Gamma
    (n/2-\alpha/2-1/2)}{\Gamma (n/2)}-
    \frac{2 \pi^{(\alpha+1)/2}}{\Gamma
    ((\alpha+1)/2)} \frac{1}{1+\alpha} (a^2 m^2)^{-n/2} \ell^{1+\alpha} \nonumber \\
    & &  _2F_1\left( \frac{n}{2},\frac{1+\alpha}{2};\frac{3+\alpha}{2};-\left(
    \frac{\ell^2}{a^2 m^2}\right) \right)
    \label{fourth_d}
    \end{eqnarray}
    Let us note that on taking the massless limit one can analytically continue the
    hypergeometric function and after straightforward calculations one gets
    \be
    -\frac{2 \pi^{(\alpha+1)/2}}{\Gamma((\alpha+1)/2)} \frac{1}{1+\alpha-n}
    \ell^{1+\alpha-n},
     \label{intadiamassless}
    \ee 
    which is of course the result one would obtain setting $m=0$ from the 
    beginning.
    Therefore all the massless singularities in (\ref{fourth_d}) correctly cancel.
    We note that in the massless limit the analytic continuation misses the UV divergencies
    stronger than the logarithmic ones.

    On again considering the case $m \ne 0$ such that
    $a(t)>H/m$ and using the result (\ref{fact}), which also holds also for $_2F_1$,
    one obtains, to the fourth adiabatic
    order, for $\langle \varphi^2 \rangle$:
    \begin{eqnarray}
    \langle \varphi^2\rangle_{(4)} &=&\frac{\hbar}{16\pi^2}  H^2
    \Bigl\{ \left[ -2+\frac{4}{3}\frac{m^2}{H^2}\right]
    \left(\frac{a m}{2 \pi^{1/2}} \right)^{d-3} \Gamma
    \left(\frac{1}{2}-\frac{d}{2}\right) +\frac{2}{9}
    \frac{m^2}{H^2}-\frac{4}{3}  + \nonumber \\
    & &
    +\frac{1}{m^2}\left[\frac{7}{45}m^2+\frac{29}{15}H^2\right]+{\cal O} \left
    ( \frac{1}{a^3} \right )\Bigr\} +{\cal O} (d-3)
    \label{fisquare_appendix}
    \end{eqnarray}


\begin{thebibliography}{99}
\bibitem{explanation} Let us remember that the evaluation of the
    renormalized expectation
    value of $\varphi^2$ is limited by the approximation of neglecting the
    infrared part of the integral in the expectation values. This
    infrared part leads to a contribution ${\cal O}(1/a^3)$ in the fourth
    order adiabatic quantities, such as in Eq. (\ref{fisquare_appendix}).
    For the adiabatic counterterms this approximation is thus very good.
    Instead, for the bare renormalized expectation
    value of $\varphi^2$, we (numerically) estimate this infrared contribution
    to be $\sim H^2/3$. Such a term is negligible with respect to the leading
    terms $\sim H^2 \log a \,, H^4/m^2$ in the final finite
    renormalized value of $\langle \varphi^2 \rangle$ (\ref{fisquare_ren}),
    but  is of the same order of magnitude of the nonleading terms
    present in Eq. (\ref{fisquare_ren}).
\bibitem{largem} We note that Eq. (\ref{desitter}) should also be correct
    for large $m$, $m > H$. In this case $\nu$ is imaginary, but the solutions
    with the Bessel functions are valid, and so is the dimensional
    regularization. From Eq. (\ref{desitter}) it is evident that the leading
    behaviour of the EMT would be ${\cal O} (m^4 g_{\mu \nu})$, which would
    correspond to a large renormalized vacuum energy. This result appears
    puzzling
    if one believes that there should not be significant
    particle production for $m \gg H$. However, this behaviour can be
    understood in the framework of adiabatic subtraction. Even for a large
    mass $m$ the Green's function of a test field in de Sitter space has
    a power-law decay, while in Minkowski space-time a massive scalar 
field
    has an exponential decay. Since the part one subtracts in the adiabatic
    subtraction is basically constructed perturbatively from the solutions for
    Minkowski space-time, we think that the subtraction is not sufficient to
    eliminate this large vacuum energy contribution.
\bibitem{senzacutoff} The renormalization of the EMT of inflaton
    fluctuations calculated without the cut-off introduced in Eq. (31) would
    lead to a result which coincides with the Bunch-Davies
    one (\ref{desitter})
    for ${\dot H} = 0$. However, this result would be
    singular since $\nu=3/2$, and the singularity due to $\psi(3/2 - \nu)$
    in the EMT persists since $m \ne 0$. As happens in the de Sitter case
    in the massless limit for minimally coupled scalar fields, the
    Bunch-Davies vacuum leads to singular expectations values. In the de
    Sitter case the two point function is singular, and for massive chaotic
    inflation both the two point function and $T_{\mu \nu}$ are singular.
    In the de Sitter case, on switching to the Allen-Folacci vacuum \cite{AF},
    one can have a regular two-point function, which however loses some
    de Sitter symmetries (it grows in time). It is interesting that we obtain
    a result which reduces to the Allen-Folacci one when $\dot{H}=0$, on just
    requiring regular expectation values. This means that our procedure with
    the cut-off is consistent with the redefinition of the zero mode for
    massless minimally coupled scalar fields, which leads to the
    Allen-Folacci vacuum.
\end{thebibliography}
    \end{document}